\documentclass{llncs}

\usepackage{latexsym}
\usepackage{amssymb}
\usepackage{amsmath}
\usepackage{psfig}
\usepackage{url}

\newcommand{\tn}{T^{\mathit{nd}}}
\newcommand{\lar}{\leftarrow}
\newcommand{\At}{\mathit{At}}
\newcommand{\GL}{\mathit{GL}}
\newcommand{\NSS}{\mathit{NSS}}
\newcommand{\lm}{\mathit{lm}}
\newcommand{\hd}{\mathit{hd}}
\newcommand{\ca}{\mathit{ca}}
\newcommand{\mca}{\mathit{mca}}
\newcommand{\bd}{\mathit{bd}}
\newcommand{\as}{\mathit{aset}}
\newcommand{\hs}{\mathit{hset}}
\newcommand{\Sup}{\mathit{Sup}}
\newcommand{\CC}{\mathit{CC}}
\newcommand{\dff}{\mathit{Def}}
\newcommand{\pth}{\mathit{path}}
\newcommand{\dom}{\mathit{Dom}}
\newcommand{\Tr}{\mathbf{t}}
\newcommand{\Fa}{\mathbf{f}}
\newcommand{\datac}{\mathrm{DC}}
\newcommand{\datan}{\mathrm{DATALOG}^{\neg}}
\newcommand{\data}{\mathrm{data}}
\newcommand{\yes}{\mathrm{YES}}
\newcommand{\con}{\mathit{Con}}
\newcommand{\cc}{\mathit{cc}}
\newcommand{\sol}{\mathit{sol}}
\newcommand{\vv}{\mathit{var}}
\newcommand{\vp}{\mathit{vpos}}
\newcommand{\size}{\mathit{size}}
\newcommand{\pred}{\mathit{pred}}
\newcommand{\pos}{\mathit{pos}}
\newcommand{\indx}{\mathit{index}}
\newcommand{\num}{\mathit{num}}
\newcommand{\bl}{\mathit{block}}
\newcommand{\inb}{\mathit{in\_block}}
\newcommand{\cl}{\mathit{cl}}
\newcommand{\vpos}{\mathit{vpos}}
\newcommand{\var}{\mathit{Var}}
\newcommand{\HU}{\mathit{HU}}
\newcommand{\pr}{\mathit{Pr}}
\newcommand{\gr}{\mathit{gr}}
\newcommand{\grnd}{\mbox{\tt{PSgrnd}}}
\newcommand{\eps}{\mbox{\tt{ePSs}}}
\newcommand{\xgrnd}{\mbox{\tt{ePSgrnd}}}
\newcommand{\at}{\mathit{At}}
\newcommand{\plc}{\mathit{PS}}
\newcommand{\eplc}{\mathit{PS}+}
\newcommand{\HB}{\mathit{HB}}
\newcommand{\plcp}{\mathit{PS}+}
\newcommand{\epl}{\mathit{ePS}}
\newcommand{\col}{\mathit{color}}
\newcommand{\gcl}{\mathit{gcl}}
\newcommand{\tc}{\mathit{tc}}
\newcommand{\vc}{\mathit{vc}}
\newcommand{\eq}{\mathit{eq}}
\newcommand{\sm}{\mathit{sum}}
\newcommand{\df}{\mathit{diff}}
\newcommand{\ltq}{\mathit{lteq}}
\newcommand{\nq}{\mathit{nq}}
\newcommand{\invc}{\mathit{invc}}
\newcommand{\hc}{\mathit{hc}}
\newcommand{\hcp}{\mathit{hc\_perm}}
\newcommand{\hce}{\mathit{hc\_edge}}
\newcommand{\start}{\mathit{start}}
\newcommand{\visit}{\mathit{visit}}
\newcommand{\clr}{\mathit{clrd}}
\newcommand{\body}{\mathit{body}}
\newcommand{\vtx}{\mathit{vtx}}
\newcommand{\edge}{\mathit{edge}}
\newcommand{\und}{{\_}}
\newcommand{\tr}{{\bf T}}
\newcommand{\fa}{{\bf F}}
\newcommand{\vph}{\varphi}
\newcommand{\rra}{\rightarrow}
\newcommand{\Ra}{\Rightarrow}
\newcommand{\La}{\Leftarrow}
\newcommand{\lla}{\leftarrow}
\newcommand{\lra}{\leftrightarrow}
\newcommand{\Lra}{\Leftrightarrow}
\newcommand{\n}{\mathbf{not}}
\newcommand{\impl}{\Leftarrow}
\newcommand{\proves}{\vdash}

\newcommand{\mc}[1]{\marginpar{\scriptsize{#1}}}
\newcommand{\vd}[1]{\text{vdW}_{#1}}

\begin{document}

\pagestyle{empty}

\title{Logic programs with monotone cardinality atoms}

\titlerunning{Logic programs with monotone cardinality atoms}

\author{Victor W. Marek\inst{1} \and Ilkka Niemel\"a\inst{2} \and
Miros\l aw Truszczy\'nski\inst{1}}
\authorrunning{Victor Marek et al.}
\institute{Department of Computer Science, University of Kentucky,\\
Lexington, KY 40506-0046, USA
\and
Department of Computer Science and Engineering\\
Helsinki University of Technology,\\
P.O.Box 5400, FIN-02015 HUT, Finland }

\maketitle

\begin{abstract}
We investigate {\em mca-programs}, that is, logic programs with clauses
built of monotone cardinality atoms of the form $kX$, where $k$ is a
non-negative integer and $X$ is a finite set of propositional atoms. We
develop a theory of mca-programs. We demonstrate that the operational
concept of the one-step provability operator generalizes to
mca-programs, but the generalization involves nondeterminism.  Our main
results show that the formalism of mca-programs is a common
generalization of (1) normal logic programming with its semantics of
models, supported models and stable models, (2) logic programming with
cardinality atoms and with the semantics of stable models, as defined by
Niemel\"a, Simons and Soininen, and (3) of disjunctive logic programming
with the {\em possible-model} semantics of Sakama and Inoue.
\end{abstract}

\section{Introduction}\label{intro}

We introduce and study logic programs whose clauses are
built of {\em monotone cardinality atoms} ({\em mc-atoms}), that is,
expressions of the form $kX$, where $k$ is a non-negative integer and
$X$ is a finite set of propositional atoms. Intuitively, $kX$ is true
in an interpretation $M$ if at least $k$ atoms in $X$ are true in
$M$.  Thus, the intended role for mc-atoms is to represent
constraints on lower bounds of cardinalities of sets. We refer
to programs with mc-atoms as {\em mca-programs}. We are motivated in
this work by the recent emergence and demonstrated effectiveness of
logic programming extended with means to model cardinality constraints
\cite{nss99,ns00,sns02}, and by the need to develop sound theoretical
basis for such formalisms.

In the paper, we develop a theory of mca-programs. In that we closely
follow the development of normal logic programming and lift all its
major concepts, techniques and results to the setting of mca-programs.
There is, however, a basic difference. Mc-atoms have, by their very
nature, a built-in nondeterminism. They can be viewed as shorthands for
certain disjunctions and, in general, there are many ways to make an
mc-atom $kX$ true.
This nondeterminism has a key consequence. The one-step provability
operator is no longer deterministic, as in normal logic programming,
where it maps interpretations to interpretations.
In the case of mca-programs, the one-step provability operator is
nondeterministic. It assigns to an interpretation $M$ a {\em set}
of interpretations, each regarded as possible
and equally likely outcomes of applying the operator to $M$.

Modulo this difference, our theory of mca-programs
parallels that of normal logic programs. First, we
introduce {\em models} and {\em supported} models of an mca-program and
describe them in terms of the one-step provability operator in much the
same way it is done in normal logic programming. To define {\em stable}
models we first define the class of {\em Horn} mca-programs by
disallowing the negation operator in the bodies of clauses. We show that
the nondeterministic one-step provability operator associates with Horn
mca-programs a notion of a (nondeterministic) computation (the
counterpart
to the bottom-up computation with normal Horn programs) and a class of
{\em derivable models} (counterparts to the least model of a normal
Horn program). We then lift the notion of the Gelfond-Lifschitz reduct
\cite{gl88} to the case of mca-programs and define a stable model of
an mca-program as a set of atoms that is a derivable model of the
reduct. A striking aspect of our construction is that all its steps
are {\em literal} extensions of the corresponding steps in the original
approach. We show that stable models behave as expected. They are
supported and, in case of Horn mca-programs, derivable.

An intended meaning of an mc-atom $1\{a\}$ is that $a$ be true. More
formally, $1\{a\}$ is true in an interpretation if and only if $a$ is
true in that interpretation. That connection implies a natural
representation of normal logic programs as mca-programs. We show that
this representation preserves all semantics we discuss in the paper.
It follows that the formalism of mca-programs can be viewed as a
direct generalization of normal logic programming.

As we noted, an extension of logic programming with direct ways to
model cardinality constraints was first proposed in
\cite{nss99}. That work defined a syntax of logic programs with 
cardinality constraints (in fact, with more general {\em weight 
constraints})
and introduced the notion of a {\em stable model}. We will refer to
programs in that formalism as {\em NSS-programs}. One of the results in
\cite{nss99} showed that NSS-programs generalized
normal logic programming with the stable-model semantics of Gelfond and
Lifschitz \cite{gl88}. However, the notion of the reduct underlying the
definition of a stable model given in \cite{nss99} is different from
that proposed by Gelfond and Lifschitz \cite{gl88} and the precise nature
of the relationship between normal logic programs and NSS-programs was
not clear.

Mca-programs explicate this relationship. We show that
the formalism of mca-programs parallels normal logic
programming. In particular, major concepts, results and techniques
in normal logic programming have counterparts in the setting of
mca-programs. We also prove that under some simple transformations,
NSS-programs are equivalent to mca-programs. Through this connection,
the theory of normal logic programming can be lifted to the setting of
NSS-programs leading to new characterizations of stable models of
NSS-programs.

Finally, we show that mca-programs not only provide an overarching
framework for both normal logic programs and NSS-programs. They are also
useful in investigating disjunctive logic programs. In the paper, we
show that logic programming with mc-atoms generalize disjunctive logic
programming with the possible-model semantics introduced in \cite{si94}.

\section{Logic programs with monotone cardinality atoms}\label{mono-ca}

Let $\At$ be a set of (propositional) {\em atoms}. An {\em mc-atom over
$\At$} (short for a {\em monotone cardinality atom over $\At$}) is any
expression of the form $k X$, where $k$ is a non-negative integer
and $X\subseteq \At$ is a {\em finite} set such that $k\leq |X|$. We call
$X$ the {\em atom set}
of an mc-atom $A= kX$ and denote it by $\as(A)$. An intuitive reading
of an mc-atom $k X$ is: {\em at least $k$ atoms in $X$ are true}.
The intended
meaning of $kX$ explains the requirement that $k \leq |X|$. Clearly,
if $k>|X|$, it is impossible to have in $X$ at least $k$ true atoms and
the expression $kX$ is equivalent to a contradiction.

An {\em mc-literal} is an
expression of the form $A$ or $\n(A)$, where $A$ is an mc-atom. An {\em
mca-clause} (short for a {\em monotone-cardinality-atom clause}) is an
expression $r$ of the form
\begin{equation}
\label{clause}
H \leftarrow L_1,\ldots, L_m,
\end{equation}
where $H$ is an mc-atom and $L_i$, $1\leq i\leq m$, are mc-literals.
We call the mc-atom $H$ the {\em head} of $r$ and denote it by
$\hd(r)$. We call the set $\{L_1,\ldots,L_m\}$ the {\em body} of $r$
and denote it by $\bd(r)$. An mca-clause is {\em Horn} if its body
does not contain literals of the form $\n(A)$. Finally, for an
mca-clause $r$, we define the {\em head set} of $r$, $\hs(r)$, by
setting $\hs(r)=\as(\hd(r))$.

Mca-clauses form {\em mca-programs}. We define the {\em head set} of
an mca-program $P$, $\hs(P)$, by $\hs(P)=\bigcup \{\hs(r)\colon r\in
P\}$ (if $P=\emptyset$,
$\hs(P)=\emptyset$, as well). If all clauses in an mca-program $P$ are
Horn, $P$ is a {\em Horn} mca-program.

One can give a declarative interpretation to mca-programs in terms of
a natural extension of the semantics of propositional logic. We say
that a set $M$ of atoms {\em satisfies} an mc-atom $k X$ if $|M\cap
X| \geq k$, and $M$ {\em satisfies} an mc-literal $\n(k X)$ if it does
not satisfy $k X$ (that is, if $|M\cap X| < k$). A set of atoms $M$
satisfies an mca-clause (\ref{clause}) if $M$ satisfies $H$ whenever $M$
satisfies all literals $L_i$, $1\leq i\leq m$. Finally, a set of atoms
$M$ satisfies an mca-program $P$ if it satisfies all clauses in $P$.
We often say ``is a model of'' instead of ``satisfies''. We use
the symbol $\models$ to denote the satisfaction relation.

The following straightforward property of mc-atoms explains the
use of the term ``monotone'' in their name.

\begin{proposition}
\label{mono}
Let $A$ be an mc-atom over a set of atoms $\At$. For every sets $M,M'
\subseteq \At$, if $M\subseteq M'$ and $M\models A$ then $M'\models
A$.
\end{proposition}

Mca-clauses also have a procedural interpretation in which they are
viewed as derivation rules. Intuitively, if an mca-clause $r$ has
its body satisfied by some set of atoms $M$, then $r$ provides {\em
support} for deriving from $M$ any set of atoms $M'$ such that
\begin{enumerate}
\item $M'$ consists of atoms mentioned in the head of $r$ ($r$ provides
no grounds for deriving atoms that do not appear in its head)
\item $M'$ satisfies the head of $r$ (since $r$ ``fires'', the
constraint imposed by its head must hold).
\end{enumerate}
Clearly, the process of deriving $M'$ from $M$ by means of $r$ is {\em
nondeterministic} in the sense that, in general, there are several sets
that are supported by $r$ and $M$.

This notion of nondeterministic derivability extends to programs and
leads to the concept of the nondeterministic one-step provability
operator. Let $P$ be an mca-program and let $M\subseteq \At$ be
a set of atoms.
We set $P(M)=\{r\in P\colon M\models \bd(r)\}$. We
call mca-clauses in $P(M)$, {\em $M$-applicable}.

\begin{definition}
Let $P$ be an mca-program and let $M\subseteq \At$. A set $M'$ is {\em
nondeterministically one-step provable} from $M$ by means of $P$, if $M'
\subseteq \hs(P(M))$ and $M'\models \hd(r)$, for every
mca-clause $r$ in $P(M)$.

The {\em nondeterministic one-step provability operator} $\tn_P$, is
a function from ${\cal P}(\At)$ to ${\cal P}({\cal P}
(\At))$ and such that for every $M \subseteq \At$, $\tn_P(M)$ consists
all sets $M'$ that are nondeterministically one-step provable from $M$
by means of $P$.
\end{definition}

As we indicate next, for every $M\subseteq \At$, $\tn_P(M)$ is
nonempty.
It follows that $\tn_P$ can be viewed as a formal representation of
a {\em nondeterministic} operator on ${\cal P}(\At)$, which assigns to
every subset $M$ of $\At$ a subset of $\At$ arbitrarily selected from
the
collection $\tn_P(M)$ of possible outcomes. Since $\tn_P(M)$ is
nonempty, this nondeterministic operator is well defined.

\begin{proposition}\label{ok}
Let $P$ be an mca-program and let $M\subseteq \At$. Then,
$\hs(P(M))\in \tn_P(M)$. In particular, $\tn_P(M)\not=\emptyset$.
\end{proposition}

The operator $\tn_P$ plays a fundamental role in our research. It allows
us to formalize procedural interpretations of mca-clauses and identify for
them matching classes of models that provide the corresponding
declarative account.

%

Our first result characterizes models of mca-programs.
This characterization is a generalization of the familiar
description of models of normal logic programs as prefixpoints of
$T_P$.

\begin{theorem}
\label{model}
Let $P$ be an mca-program and let $M\subseteq \At$. The set $M$ is a
model of $P$ if and only if there is $M'\in \tn_P(M)$ such that
$M'\subseteq M$.
\end{theorem}
%

A straightforward corollary states that every
mca-program has a model.

\begin{corollary}
Let $P$ be an mca-program. Then, $\hs(P)$ is a model of $P$.
\end{corollary}

Models of mca-programs may contain elements that have no support in a
program and the model itself. For instance, let us consider an
mca-program $P$ consisting of the clause:
$1\{p,q\} \lar \n(1\{q\})$,
where $p$ and $q$ are two different atoms. Let $M_1=\{q\}$. Clearly,
$M_1$ is a model of $P$. However, $M_1$ has no support in $P$ and
itself. Indeed, $\tn_P(M_1)=\{\emptyset\}$ and so, $P$ and $M_1$ do not
provide support for any atom. Similarly, another model of $P$, the set
$M_2=\{p,r\}$, where $r \in \At$ is an atom different from $p$ and $q$,
has no support in $P$ and itself. We have $\tn_P(M_2) = \{\{p\},\{q\},
\{p,q\}\}$ and so, $p$ has support in $P$ and $M_2$, but $r$ does not.
Finally, the set $M_3=\{p\}$, which is also a model of $P$, {\em has
support}
in $P$ and itself. Indeed, $\tn_P(M_3)=\{\{p\},\{q\},\{p,q\}\}$ and
there is a way to derive $M_3$ from $P$ and $M_3$. We
formalize now this discussion in the following definition.

\begin{definition}\label{supp}
Let $P$ be an mca-program. A set of atoms $M$ is a {\em supported
model} of $P$ if $M\in \tn_P(M)$.
\end{definition}

The use of the term ``model'' is justified. By Theorem \ref{model},
supported models of $P$ are indeed models of $P$, as stated in the
following result.

\begin{corollary}
Every supported model of an mca-program $P$ is a model of $P$.
\end{corollary}

Finally, we have the following characterization of supported models.

\begin{proposition}\label{ch-sup}
Let $P$ be an mca-program. A set $M\subseteq\At$ is a supported model
of $P$ if and only if $M$ is a model of $P$ and $M\subseteq\hs(P(M))$.
\end{proposition}

\section{Horn mca-programs}\label{horn}

To introduce {\em stable} models of mca-programs, we need first to
study Horn mca-programs.
With each Horn mca-program $P$ one can associate the concept of a {\em
$P$-computation}. Namely, a {\em $P$-computation} is a sequence
$(X_n)_{n=0,1,\ldots}$ such that $X_0=\emptyset$ and, for every
non-negative integer $n$,
\begin{enumerate}
\item $X_n\subseteq X_{n+1}$, and
\item $X_{n+1}\in \tn_P(X_{n})$.
\end{enumerate}
Given a computation $t=(X_n)_{n=0,1,\ldots}$, we call
$\bigcup_{n=0}^{\infty} X_n$ the {\em result} of the computation $t$
and denote it by $R_t$.

\begin{proposition}\label{head}
Let $P$ be a Horn mca-program and let $t$ be a $P$-computation.
Then $R_t\subseteq \hs(P(R_t))$.
\end{proposition}

If $P$ is a Horn mca-program then
$P$-computations
exist. Let $M$ be a model of $P$. We define the
sequence $t^{P,M}=(X^{P,M}_n)_{n=0,1,\ldots}$ as follows. We set
$X^{P,M}_0=\emptyset$ and, for every $n\geq 0$, $X^{P,M}_{n+1} =
\hs(P(X^{P,M}_n))\cap M$.

\begin{theorem}\label{comp}
Let $P$ be a Horn mca-program and let $M\subseteq \At$ be its model.
The sequence $t^{P,M}$ is a $P$-computation.
\end{theorem}
%

We call the $P$-computation $t^{P,M}$ the {\em canonical}
$P$-computation for $M$. Since every mca-program $P$ has models,
we obtain the following corollary.

\begin{corollary}
Every Horn mca-program has at least one computation.
\end{corollary}

The results of computations are supported
models (and, thus, also models) of Horn mca-programs.

\begin{proposition} \label{p.model}
Let $P$ be a Horn mca-program and let $t$ be a $P$-computation. Then,
the result of $t$, $R_t$, is a supported model of $P$.
\end{proposition}

We use the concept of a computation to identify a certain class of
models of Horn mca-programs.

\begin{definition}
Let $P$ be a Horn mca-program. We say that a set of atoms $M$
is a {\em derivable model} of $P$ if there exists a $P$-computation
$t$ such that $M = R_t$.
\end{definition}

Derivable models can be obtained as results of their own canonical
computations.

\begin{proposition}
Let $M$ be a derivable model of a Horn mca-program $P$. Then
$M=R_{t^{P,M}}$.
\end{proposition}

Proposition \ref{p.model} and Theorem \ref{comp} entail several
properties of Horn mca-programs, their computations and models. We
gather them in the following corollary.

\begin{corollary} \label{p.sum}
Let $P$ be a Horn mca-program. Then:
\begin{enumerate}
\item $P$ has at least one derivable model.
\item $P$ has a largest derivable model.
\item Every derivable model of $P$ is a supported model of $P$.
\item For every model $M$ of $P$ there is a derivable model $M'$ of $P$
such that $M'\subseteq M$.
\item Every minimal model of $P$ is derivable.
\end{enumerate}
\end{corollary}
%
%
%
%

\section{Stable models of mca-programs}\label{stable}

We will now use the results of the two previous sections to introduce and
study the class of {\em stable} models of mca-programs.

\begin{definition}\label{d-stable}
Let $P$ be an mca-program and let $M\subseteq \At$. The {\em reduct}
of $P$ with respect to $M$, $P^M$ in symbols, is a Horn mca-program
obtained from $P$ by (1) removing from $P$ every clause containing in
the body a literal $\n(A)$ such that $M\models A$, and (2) removing all
literals of the form $\n(A)$ from all remaining clauses in $P$.
A set of atoms $M$ is a {\em stable} model of $P$ if $M$ is a derivable
model of the reduct $P^M$.
\end{definition}

Stable models of an mca-program $P$ are indeed models of $P$. Thus, the
use of the term ``model'' in their name is justified. In fact, a
stronger property holds: stable models of mca-programs are supported.

\begin{proposition}\label{m-stable}
Let $P$ be an mca-program. If $M\subseteq \At$ is a stable model of $P$
then $M$ is a supported model of $P$.
\end{proposition}

With the notion of a stable model in hand, we can strengthen
Proposition \ref{p.model}.

\begin{proposition} \label{der-stb}
Let $P$ be a Horn mca-program. A set of atoms $M\subseteq \At$ is a
derivable model of $P$ if and only if $M$ is a stable model of $P$.
\end{proposition}

We will now describe a procedural characterization of stable models of
mca-programs, relying on a notion of a computation related to but
different from
the one we discussed in Section \ref{horn} in the context of Horn
programs. A difference is that now at each stage in a computation we
must make sure that once a clause is applied, it remains applicable at
{\em any} stage of the process. It is not {\em a priori} guaranteed due
to the presence of negation in the bodies of general mca-clauses.

A formal definition is as follows. Let $P$ be an mca-program. A
sequence $\varepsilon = (X_n
)_{n = 0, 1,2,\ldots}$ is a {\em quasi $P$-computation},
if $X_0 = \emptyset$
and if for every $n=0,1,\ldots$ there is a clause $r_n\in P$ such that
\begin{enumerate}
\item $X_n \models \bd (r_n)$.
\item there is $X\subseteq \hs(r_n)$ such that $X\models\hd(r_n)$ and
$X_{n+1} = X_n \cup X$ (this $X$ is what is ``computed'' by applying
$r_n$).
\item for every $i=0,1\ldots,n$ and for every mc-atom $kX$ occurring
negated in $\bd(r_i)$, $X_{n+1} \not\models kX$.
\end{enumerate}
We call the set $\bigcup_{1\leq k< \omega} X_k$ the {\em result} of the
quasi $P$-computation $\varepsilon$.

\begin{theorem} \label{sta-comp}
A set of atoms $M$ is a stable model of $P$ if and only if $M$ is a
model of $P$ and for some quasi
$P$-computation $\varepsilon$, $M$ is the result
of $\varepsilon$.
\end{theorem}

Theorem \ref{sta-comp} states that if we apply clauses {\em carefully},
making sure that at no stage we satisfy an mc-atom appearing negated
in clauses applied so far (including the one selected to apply
at the present stage) and we ever compute a model in this way, then
this model is a stable model of $P$. Conversely, every stable model
can be obtained as a result of such a {\em careful} computation.

%

\section{Extension of mca-programs by constraint mca-clauses}
\label{constraints}

We can extend the language of mca-programs by allowing clauses with
the empty head. Namely, we define a {\em constraint mca-clause} to be
an  expression $r$ of the form
\begin{equation}
\label{cl}
\leftarrow L_1,\ldots, L_m,
\end{equation}
where $L_i$, $1\leq i\leq m$, are mc-literals.

The notion of satisfiability that we introduced for mca-clauses extends
to the case of mca-constraints. A set of atoms $M$ {\em satisfies} a
constraint $r$ if there is a literal $L\in \bd(r)$ such that $M\not
\models L$. We can now extend the definitions of supported and stable
models to the more general class of mca-programs with constraint
mca-clauses as follows.

\begin{definition}
Let $P$ be an mca-program with constraint mca-clauses. A set of atoms
$M$ is a {\em supported} ({\em stable}) model of $P$ if $M$ is a
supported (stable) model of $P'$, where $P'$ consists of all
non-constraint mca-clauses in $P$, and if $M$ is a model of all
constraint mca-clauses in $P$.
\end{definition}

Let us observe that several of our earlier results such as Proposition
\ref{m-stable} and Theorem \ref{sta-comp} lift {\em verbatim} to the
case of programs with constraints.

\section{Mca-programs and normal logic programming}\label{nlp-mca}

An mc-atom $1\{a\}$ is true in a model $M$ if and only if $a$ is true
in $M$. Thus, intuitively, $1\{a\}$ and $a$ are equivalent. That
suggests a way to interpret normal clauses and programs as mca-clauses
and mca-programs. Let 
\[
r=\ \ \ c \lar a_1,\ldots,a_m,\n(b_1),\ldots,\n(b_n).
\]
By $\mca(r)$ we mean the mc-clause
\[
1\{c\} \lar 1\{a_1\},\ldots,1\{a_m\},\n(1\{b_1\}),\ldots,\n(1\{b_n\}).
\]
(If all $a_i$ and all $b_i$ are distinct, which we can assume
without loss of generality, a simpler translation, $1\{c\} \lar
m\{a_1,\ldots,a_m\},\n(1\{b_1,\ldots,b_n\})$, could be used.)
Moreover, given a normal program $P$, we set
$\mca(P)=\{mc(r)\colon r\in P\}$.

This encoding interprets normal logic programs as mca-programs so that
basic properties and concepts of normal logic programming can be viewed
as special cases of properties and concepts in
mca-programming. In the following theorem, we gather several results
establishing appropriate correspondences.

\begin{theorem}
\label{lp-mca}
Let $P$ be a normal logic program and let $M$ be a set of atoms.
\begin{enumerate}
\item $P$ is a Horn program if and only if $\mca(P)$ is a Horn
mca-program.
\item If $P$ is a Horn program then the least model of $P$
is the only derivable model of $\mca(P)$.
\item $\{T_P(M)\}=\tn_{\mca(P)}(M)$.
\item $\mca(P^M) = \mca(P)^M$.
\item $M$ is a model (supported model, stable model) of $P$ if and
only if $M$ is a model (supported model, stable model) of $\mca(P)$.
\end{enumerate}
\end{theorem}

Finally, we identify a class of mca-programs, which offers a most
direct generalization of normal logic programming.

\begin{definition}
\label{det}
An mca-clause $r$ is {\em deterministic} if $\hd(r) =1\{a\}$, for some
atom $a$. An mca-program is {\em deterministic} if every clause in $P$
is deterministic.
\end{definition}

The intuition behind the term is clear. If the head of an mca-clause is
of the form $1\{a\}$, then there is only one possible effect of
applying the clause: $a$ has to be concluded. Thus, the nondeterminism that
arises in the context of arbitrary mc-atoms disappears. Formally, we
capture this property in the following result.

\begin{proposition}
\label{det1}
Let $P$ be a deterministic mca-program. Then, for every set of atoms
$M$, $\tn_P(M)=\{M'\}$, for some set of atoms $M'$.
\end{proposition}

Thus, for a deterministic mca-program $P$, the
operator $\tn_P$ is deterministic and, so, can be regarded as an
operator with both the domain and codomain ${\cal P}(\At)$. We will
write $T^d_P$, to denote it. Models, supported models and
stable models of a deterministic mca-program can be introduced in
terms of the operator $T^d_P$ in exactly the same way the corresponding
concepts are defined in normal logic programming. In particular, the
algebraic treatment of logic programming developed in
\cite{fi99,przy90,dmt00a} applies literally to deterministic
mca-programs and results in a natural and direct extension of normal
logic programming.
We will explicitly mention just one result here that will be of
importance later in the paper.

\begin{proposition}
\label{horn-det}
Let $P$ be a deterministic Horn program. Then $P$ has exactly one
derivable model and this model is the least model of $P$.
\end{proposition}

\section{Mca-programs and NSS-programs}
\label{mca-cc}

We will first briefly review the concept of an NSS-program
\cite{nss99}, the semantics of stable models of such programs,
as introduced in \cite{nss99}, and then relate this formalism to that
of mca-programs.

A {\em cardinality atom} (c-atom, for short) is an expression of
the form $kXl$, where $X\subseteq \At$, and $l$ and $k$ are integers
such that $0 \leq k \leq l\leq |X|$. We call $X$ an {\em atom set} of
a c-atom $A=kXl$ and, as before, we denote it by
$\as(A)$\footnote{To be precise, \cite{nss99} allows also for negated
atoms to appear as elements of $X$. One can eliminate occurrences of
negative literals by introducing new atoms. Thus, for this work,
we decided to restrict the syntax of NSS-programs.}.

We say that a set of atoms $M$ satisfies a c-atom $kXl$ if $k\leq
|M\cap X|\leq l$ ($M\models kXl$, in symbols). It is clear that when
$k=0$ or $l=|X|$, the corresponding inequality is trivially true. Thus,
we omit from the notation $k$, if equal to 0, and $l$, if equal to
$|X|$.

A {\em cardinality-atom clause} (ca-clause, for short) is an expression
$r$ of the form
\[
 A \leftarrow B_1,\ldots, B_n,
\]
where $A$ and $B_i$, $1\leq i\leq n$, are c-atoms. We call $A$ the head
of $r$ and $\{B_1,\ldots,B_n\}$ the {\em body} of $r$. We denote them
by $\hd(r)$ and $\bd(r)$, respectively. A {\em ca-program} is a
collection of ca-clauses.

We say that a set $M\subseteq \At$ {\em satisfies} a ca-clause $r$ if
$M$ satisfies $\hd(r)$ whenever it satisfies each c-atom in the body of
$r$. We say that $M$ satisfies a ca-program $P$ if $M$ satisfies each
ca-clause in $P$. We write $M\models r$ and $M\models P$ in these
cases, respectively.

We will now recall the concept of a stable model of a ca-program
\cite{nss99}. Let $P$ be an NSS-program and let $M\subseteq \At$.
By the {\em NSS-reduct} of $P$ with respect to $M$ we mean the
NSS-program obtained by:
\begin{enumerate}
\item eliminating from $P$ every clause $r$ such that $M\not\models B$,
for at least one c-atom $B\in \bd(r)$.
\item replacing each remaining ca-clause $r= kXl \lar k_1Y_1l_1,\ldots
k_nY_nl_n$ with all clauses of the form
$1\{a\} \lar k_1Y_1,\ldots,k_nY_n$,
where $a \in X \cap M$.
\end{enumerate}
With some abuse of notation, we denote the resulting program by $P^M$
(the type of the program determines which reduct we have in mind).
It is clear that $P^M$ is a deterministic Horn mca-program. Thus,
it has a least model, $\lm(P^M)$.

\begin{definition}
\label{def-nss}
Let $P$ be a ca-program. A set $M\subseteq \At$ is a {\em stable model}
of $P$ if  $M=\lm(P^M)$ and $M\models P$.
\end{definition}

We will now show that the formalisms of mca-programs and ca-programs
with their corresponding stable-model semantics are equivalent. We
start by describing an encoding of ca-clauses and ca-programs by
mca-clauses and mca-programs. To simplify the description of the
encoding
and make it uniform, we assume that all bounds are present (we recall
that whenever any of the bounds are missing from the
notation, they can be introduced back). Let $r$ be the following
ca-clause:
$k X l \lar k_1 X_1 l_1,\ldots, k_m X_m l_m.$
We represent this ca-clause by a pair of mca-clauses, $e_{\mca}^1(r)$
and $e_{\mca}^2(r)$ that we define as the following two mca-clauses,
respectively:
\[
k X  \lar k_1 X_1, \ldots, k_m X_m, \n((l_1+1)X_1), \ldots, \n((l_m+1)X_m),
\]
and
\[
\lar (l+1)X, k_1 X_1, \ldots, k_m X_m, \n((l_1+1)X_1),
\ldots, \n((l_m+1)X_m).
\]
Given a ca-program $P$, we translate it into an mca-program
\[
e_{\mca}(P)= \bigcup_{r \in P}
\{e^1_{\mca}(r), e^2_{\mca}(r)\}.
\]

\begin{theorem}
\label{ca-mca}
Let $P$ be a ca-program. A set of atoms $M$ is a stable model of $P$,
as defined for ca-programs, if and only if $M$ is a stable model of
$e_{\mca}(P)$, as defined for mca-programs.
\end{theorem}

This theorem shows that the formalism of mca-programs is at least as
expressive as that of ca-programs. The converse
is true as well: ca-programs are at least as expressive as
mca-programs. Let $r$ be the following mca-clause:
\[
kX \lar k_1X_1,\ldots, k_mX_m ,\n(l_1 Y_1), \ldots, \n(l_nX_n).
\]
We define $e_{\ca}(r)$ as follows. If there is $i$, $1\leq i\leq n$,
such that $l_i=0$, we set $e_{\ca}(r)=\ \ kX\lar kX$ (in fact any
tautology would do). Otherwise, we set
\[
e_{\ca}(r) = \ \ kX \lar k_1X_1,\ldots, k_mX_m, Y_1 (l_1-1),\ldots,
Y_n(l_n-1).
\]
Given an mca-program $P$, we define
$e_{\ca}(P)=\{e_{\ca}(r)\colon r\in P\}$.

\begin{theorem}
\label{mca-ca}
Let $P$ be an mca-program. A set of atoms $M$ is a stable model of $P$,
as defined for mca-programs, if and only if $M$ is a stable model of
$e_{\ca}(P)$, as defined for ca-programs.
\end{theorem}

Theorems \ref{ca-mca} and \ref{mca-ca} establish the equivalence of
ca-programs and mca-programs with respect to the stable
model semantics. The same translations also preserve the concept of a
model. Finally, Theorem \ref{ca-mca} suggests a way to introduce
the notion of a supported model for a ca-program: a set of atoms $M$ is
defined to be a {\em supported} model of a ca-program $P$ if it is a
supported model of the mca-program $e_{\mca}(P)$. With this definition,
the two translations $e_{mca}$ and $e_{\ca}$ also preserve the concept
of a supported model.

We also note that this equivalence demonstrates that ca-programs
with the semantics of stable models as defined in \cite{nss99} can be
viewed as a generalization of normal logic programming. It follows from
Theorems \ref{lp-mca} and \ref{mca-ca} that the encoding of normal logic
programs as ca-programs, defined as the composition of the translations
$\mca$ and $e_{\ca}$, preserves the semantics of models, supported
models and stable models (an alternative proof of this fact, restricted
to the case of stable models only was first given in \cite{nss99} and
served as a motivation for the class of ca-programs and its stable-model
semantics). This result is important, as it is not at all evident that
the NSS-reduct and Definition \ref{def-nss} generalize
the semantics of stable models as defined in \cite{gl88}.

Given that the formalisms of ca-atoms and mca-atoms are equivalent,
it is important to stress what differs them. The advantage of
the formalism of ca-programs is that it does not require the negation
operator in the language. The strength of the formalism of mca-programs
lies in the fact that its syntax so closely resembles that of normal
logic programs, and that the development of the theory of mca-programs
so closely follows that of the normal logic programming.

\section{Mca-programs and disjunctive logic programs}\label{mca-dp}

The formalism of mca-programs also extends an approach to disjunctive
logic
programming, proposed in \cite{si94}. In that paper, the authors
introduced and investigated a semantics of {\em possible models} for
disjunctive logic programs. We will now show that disjunctive
programming with the semantics of possible models is a special case of
the logic mca-programs with the semantics of stable models.

Let $r$ be a disjunctive logic program clause of the form:
\[
c_1\vee\ldots\vee c_k \lar a_1,\ldots,a_m,\n(b_1),\ldots,\n(b_n),
\]
where all $a_i$, $b_i$ and $c_i$ are atoms. We define an mca-clause
\[
\mca_d(r)=\ \ 1\{c_1,\dots,c_k\} \lar 1\{a_1\},\ldots,1\{a_m\},
            \n(1\{b_1\}),\ldots,\n(1\{b_n\}).
\]
For a disjunctive logic program $P$, we define $\mca_d(P)=\{\mca_d(r)
\colon r\in P\}$. We have the following theorem.

\begin{theorem}
Let $P$ be a disjunctive logic program. A set of atoms $M$ is a
possible model of $P$ if and only if $M$ is a stable model of the
mca-program $\mca_d(P)$.
\end{theorem}

We also note that there are strong analogies between the approach we
propose here and some of the techniques discussed in \cite{si94}. In
particular, \cite{si94} presents a computational procedure for
disjunctive programs without negation that is equivalent to our notion
of a $P$-computation. We stress however, that the class of mca-programs
is more general and that our approach, consistently exploiting properties
of an operator $\tn_P$, is better aligned with a standard development of
normal logic programming.

\section{Discussion}\label{disc}

Results of our paper point to a central position of mca-programs among
other logic programming formalisms. First, mca-programs form a natural
generalization of normal logic programs, with most concepts and
techniques closely patterned after their counterparts in normal logic
programming. Second, mca-programs with the stable-model semantics
generalize disjunctive logic programming with the possible-model
semantics of \cite{si94}.
Third, mca-programs provide direct means to model cardinality
constraints, a feature that has become broadly recognized as essential
to computational knowledge representation formalisms. Moreover, it turns
out that mca-programs are, in a certain sense that we made precise in
the paper, equivalent, to logic programs with cardinality atoms
proposed and studied in \cite{nss99}. Thus, mca-programs provide a
natural link between normal logic programs and the formalism of
\cite{nss99}, and help explain the nature of this relationship, hidden
by the original definitions in \cite{nss99}.

In this paper, we outlined only the rudiments of the theory of
mca-programs. There are several questions that follow from
our work and that deserve more attention. First, our
theory can be extended to the case of programs built of {\em
monotone-weight atoms}, that is, expressions of the form $a \{p_1:w_1,
\ldots,p_k:w_k\}$, where $a$, $w_1,\ldots w_k$ are non-negative reals
and $p_1,\ldots,p_k$ are propositional atoms. Intuitively, such an atom
is satisfied by an interpretation (set of atoms) $M$ if the sum of
weights assigned to atoms in $M\cap \{p_1,\ldots,p_k\}$ is at least
$a$.

Next, there is a question whether Fages lemma \cite{fag94} generalizes
to mca-programs. If so, for some classes of programs, one could reduce
stable-model computation to satisfiability checking for propositional
theories with cardinality atoms \cite{et01a,lt03}. That, in turn,
might lead to effective computational methods, alternative to direct
algorithms such as {\em smodels} \cite{ns96} and similar in spirit to
the approach of {\em cmodels} \cite{el03,cmodels}.

Another interesting aspect concerns some syntactic modifications and
``normal form representations'' for mca-programs. For instance, at a
cost of introducing new atoms, one can rewrite any mca-program into
a {\em simple} mca-program in which every mca-clause contains at most
one mca-literal in its body and in which the use of negation is
restricted (but not eliminated). We will present these results in a
full version of the paper.

The emergence of a nondeterministic one-step provability operator is
particularly intriguing. It suggests that, as in the case of normal
logic programming \cite{fi99,przy90}, the theory of mca-programs can be
developed by algebraic means. For that to happen, one would need
techniques for handling nondeterministic operators on lattices, similar
to those presented in the deterministic operators in \cite{dmt00a,dmt02}.
That approach might ultimately lead to a generalization of the
well-founded semantics to the case of mca-programs.


\section*{Acknowledgments}

The second author was supported by the Academy of Finland grant 53695.
The other two authors were supported by the NSF grants IIS-0097278 and
IIS-0325063.


\end{document}